\documentclass[aps,physrev,reprint,superscriptaddress,longbibliography,nofootinbib]{revtex4-2}

\usepackage{siunitx}
\usepackage{physics}
\usepackage[version=4]{mhchem}
\usepackage{graphicx}
\usepackage{amsmath}

\begin{document}

\title{Autonomous Reshaping of Expression Landscapes by DNA Methylation}

\author{Kaifeng Wang}
\affiliation{Center for Life Sciences, Academy for Advanced Interdisciplinary Studies, Peking University, Beijing, China}

\author{Ming Han}
\email{hanmingcr@pku.edu.cn}
\affiliation{Center for Life Sciences, Academy for Advanced Interdisciplinary Studies, Peking University, Beijing, China}
\affiliation{Center for Quantitative Biology, Academy for Advanced Interdisciplinary Studies, Peking University, Beijing, China}
\affiliation{Department of Physics, Peking University, Beijing, China}

\begin{abstract}
DNA methylation is usually treated as an epigenetic memory mark: transcriptional history is
written into regulatory DNA and later stabilizes a chosen cell identity.
This picture explains persistence, but it makes memory passive.
Here we show that the same promoter-level coupling required for methylation memory can instead
turn methylation into an internal control variable for regulatory dynamics.
Transcription-factor occupancy protects regulatory DNA from methylation, while methylation shifts
later transcription-factor binding thresholds.
Under time-scale separation, this reciprocal loop separates into fast expression dynamics
conditioned on methylation and a slow methylation flow written by expression.
Minimal promoter, self-activation, and fate-toggle models show that this feedback does more than
preserve a past state: it autonomously reshapes the expression landscape.
In a methylation-coupled toggle, the preferred expression state can move continuously through
single-well drift, allowing commitment without first entering a multiwell regime.
Stochastic simulations further show that evolving methylation reduces fate reversals relative to a
frozen landscape, making weak early expression bias more predictive of later fate.
These results recast DNA methylation from a downstream stabilizer of cell identity into a slow
dynamical coordinate that can help determine how regulatory states are chosen.
\end{abstract}

\maketitle


\section*{Introduction}

DNA methylation is commonly treated as an epigenetic memory mark: transcriptional history is
recorded in methylation, which later stabilizes the resulting regulatory state and cell identity
\cite{bird2002dna,li2014dna,baylin2005dna,fan2024promoter}.
This view reduces methylation-transcription coupling to a temporal sequence in which expression
first chooses a state and methylation preserves it.
The sequence explains why differentiated states persist, but it hides the dynamical premise on
which memory depends.
For methylation to record transcriptional history, expression must write it; for that memory to
matter, methylation must be read by future expression.

\indent
Recent studies reveal a potential molecular mechanism for this write-read loop.
Transcription factor occupancy and methylation machinery can compete at promoters and other
regulatory DNA: bound transcription factors (TFs) can protect CpG-rich regions from de novo
methylation or favor demethylation
\cite{brandeis1994sp1,macleod1994sp1,stadler2011dna,feldmann2013transcription,domcke2015competition,fan2024promoter},
while methylation can reduce later transcription factor binding or shift effective binding
thresholds
\cite{watt1988cytosine,iguchi1989cpg,yin2017impact,hernandezcorchado2022base,domcke2015competition,chen2021mathematical}.
This competition places writing and reading in the same local circuit, rather than in
separate developmental stages.

\indent
Related theory has begun to treat epigenetic state as a dynamical partner of gene regulation.
DNA-methylation models show that methylation can shift stability boundaries and basins of
attraction \cite{chen2021mathematical}, while chromatin-state and slow-promoter models show that
slow regulatory variables can reshape probability landscapes, generate hysteresis or
homeorhesis, and stabilize memory
\cite{bhattacharyya2020stochastic,sood2021quantifying,bruno2022epigenetic,matsushita2020homeorhesis,alradhawi2019multimodality}.
Beyond shifting fate stability and basin structure, how does promoter-level
methylation--transcription coupling change regulatory dynamics?

We answer this question with a sequence of minimal models.
A single regulatory locus, motivated by recent promoter-memory experiments
\cite{fan2024promoter}, first isolates residual methylation as the slow coordinate left by
promoter activity.
Adding self-activation then reveals how this memory coordinate can become an internal control
variable for the evolution of expression landscape.
Finally, in a fate toggle, the same loop produces single-well drift: a commitment mode in which
the preferred expression state moves continuously, rather than emerging through a classical
bifurcation of a Waddington-like landscape
\cite{kim2007potential,huang2007bifurcation,ferrell2012bistability,saez2022dynamical}.
Together, these examples show that methylation not only preserves a chosen fate, but also
affects fate decisions by reshaping regulatory dynamics.

\section*{Results}

\begin{figure*}[t!]
  \centering
  \includegraphics[width=\textwidth]{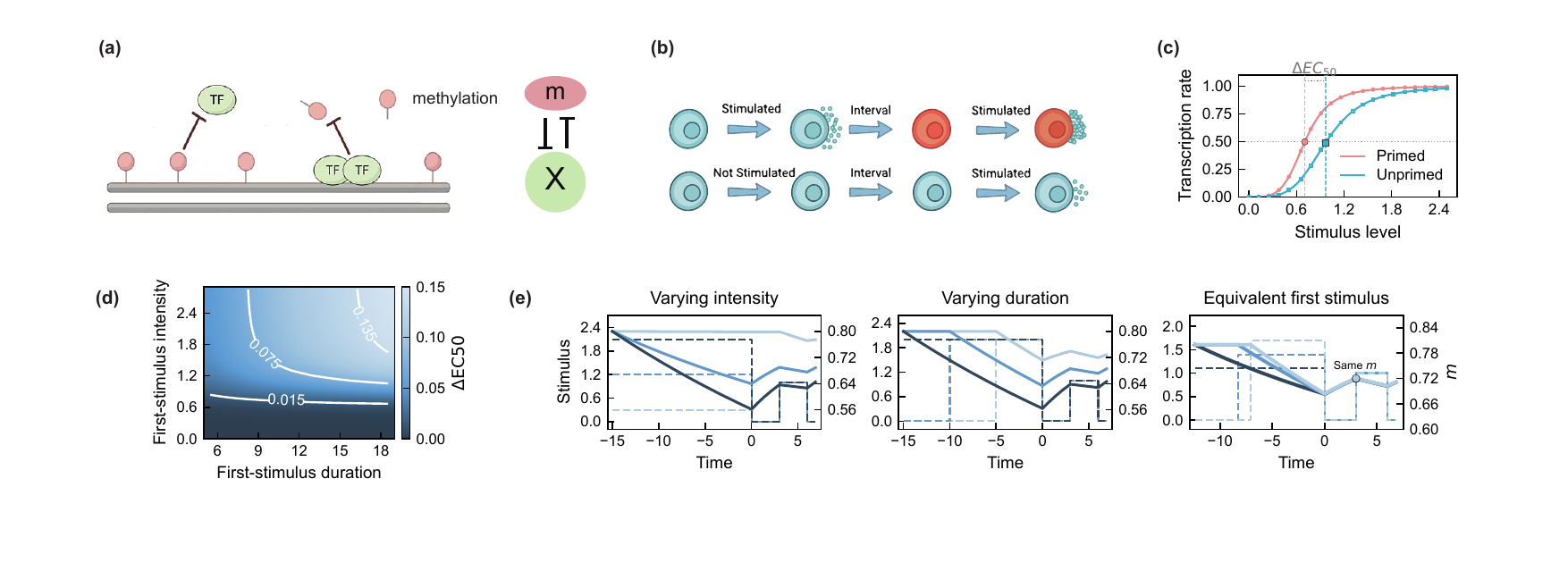}
  \caption{\textbf{Residual methylation is the memory coordinate of the minimal feedback loop.}
    \textbf{(a)}~TF--DNMT competition couples fast TF occupancy to slow promoter methylation.
    \textbf{(b)}~A two-stimulus protocol separates priming from the second response.
    \textbf{(c)}~Priming shifts the second dose response relative to an unprimed locus.
    \textbf{(d)}~The second-response threshold shift increases with first-stimulus intensity
    and duration.
    \textbf{(e)}~Different priming histories that leave the same residual methylation deviation
    $\Delta m$ produce the same second-response threshold shift, showing that methylation
    compresses transcriptional history into a single slow coordinate.}
  \label{fig:fig1}
\end{figure*}

\noindent
We first strip the coupling down to a single regulatory locus.
Consider a CpG-rich promoter exposed to a transcription factor or inducer with effective
concentration $x$ (Fig.~\ref{fig:fig1}a).
Transcription factors and methylation enzymes compete at the regulatory sequence
\cite{chen2021mathematical,domcke2015competition,heberle2019sensitivity,kribelbauer2020towards,lemma2022pioneer}.
In this local circuit, TF occupancy protects the promoter from methylation, while methylation
raises the threshold for later occupancy.
The promoter state is therefore written by past occupancy and read by future occupancy.

The readout is transcription factor binding.
Fast binding gives the occupancy
\begin{equation}
  h(x,m) = \frac{x^n}{x^n + K^n(m)},
\end{equation}
where $n=4$ is the cooperativity constant, $m\in[0,1]$ is the fraction of promoter
CpGs that are methylated, and $K(m)$ is the methylation-dependent effective dissociation
constant, equivalently the half-saturation concentration.
Methylation changes TF occupancy by changing binding free energy
\cite{stormo1998specificity,zuo2017measuring,kribelbauer2017quantitative}.
Rather than model each CpG separately, we summarize the effect of methylation by a first-order
expansion of the binding energy around the resting level $m_0$.
\begin{equation}
  K(m) = \exp\!\bigl(-\beta\,(\varepsilon_0 + \varepsilon_1(m - m_0))\bigr).
\end{equation}
Here $\beta = 1/(k_B T)$ is the Boltzmann factor and $\varepsilon_1<0$, so that increasing
methylation raises $K(m)$ and shifts the dose response to higher TF concentration.

The write arm is promoter methylation dynamics:
\begin{equation}
  \frac{dm}{dt} = -r_m\,m + p_m(1-m)\bigl(1 - h(x,m)\bigr).
  \label{eq:methylation-dynamics}
\end{equation}
Here $r_m$ and $p_m$ are the demethylation and methylation rates, respectively.
The factor $(1-h)$ implements promoter protection: methylation enzymes act most effectively when
the promoter is unoccupied, and are excluded as TF occupancy increases.
Together, the read and write arms close the minimal promoter loop.

This minimal loop gives a theoretical interpretation of recent promoter-memory experiments in
mammalian cells, where two sequential stimuli were used to separate priming from the later
response \cite{fan2024promoter}.
A first stimulus primes the locus, a stimulus-free interval removes the immediate occupancy
response, and a second stimulus measures the shifted dose response (Fig.~\ref{fig:fig1}b).
Priming shifts the second dose response relative to the unprimed locus
(Fig.~\ref{fig:fig1}c), and the shift grows with the strength and duration of the first stimulus
(Fig.~\ref{fig:fig1}d).
The collapse in Fig.~\ref{fig:fig1}e identifies methylation as a sufficient memory variable:
different intensity--duration histories produce the same shift in the second dose response
whenever they leave the same residual methylation deviation $\Delta m$.
Residual methylation therefore compresses past regulatory activity into the variable that
controls the next promoter response.

\begin{figure*}[t!]
  \centering
  \includegraphics[width=\textwidth]{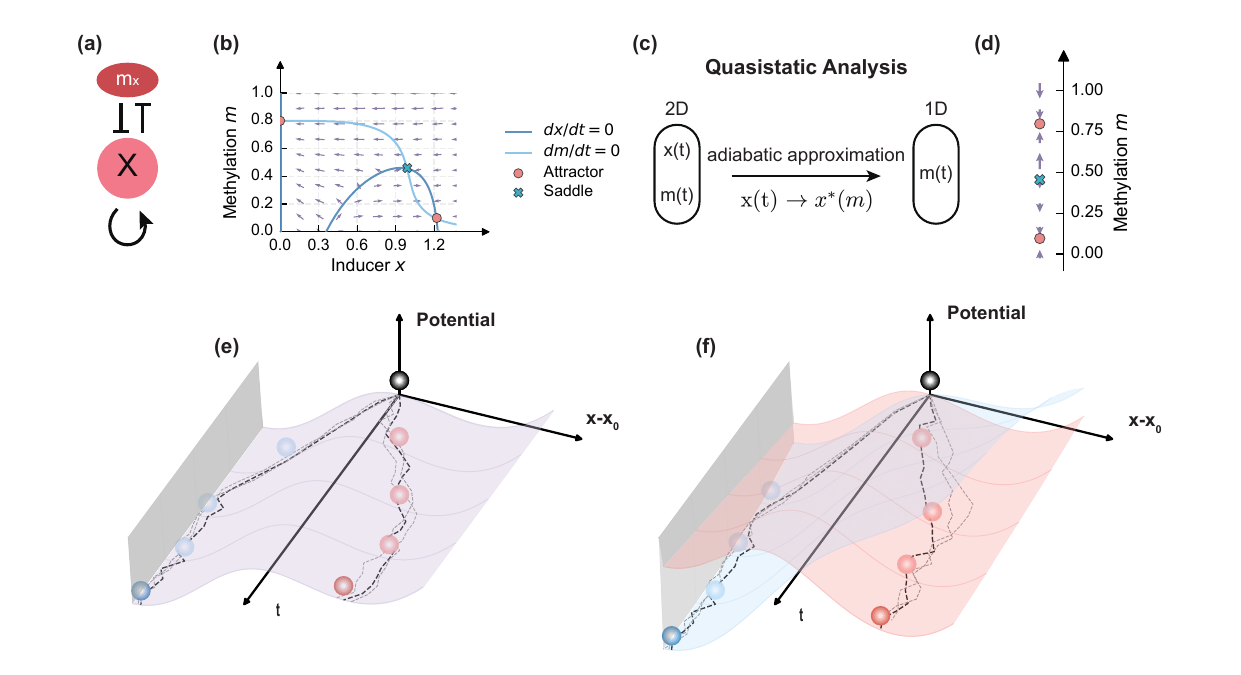}
  \caption{\textbf{Self-activation converts methylation memory into landscape motion.}
    \textbf{(a)}~A self-activating gene closes the feedback between expression and local
    methylation.
    \textbf{(b)}~At fixed methylation, nullclines and fixed points show that methylation shifts
    the fast expression dynamics.
    \textbf{(c)}~Time-scale separation lets expression equilibrate on a stable branch
    $x^\ast(m)$.
    \textbf{(d)}~Evaluating methylation dynamics on this branch gives a closed slow flow for
    $m$.
    \textbf{(e)}~With $m$ fixed, the expression landscape is frozen and does not respond to
    history.
    \textbf{(f)}~With feedback active, methylation evolves and autonomously tilts the potential
    $U(x;m)$.}
  \label{fig:fig2}
\end{figure*}

\begin{figure*}[t!]
  \centering
  \includegraphics[width=\textwidth]{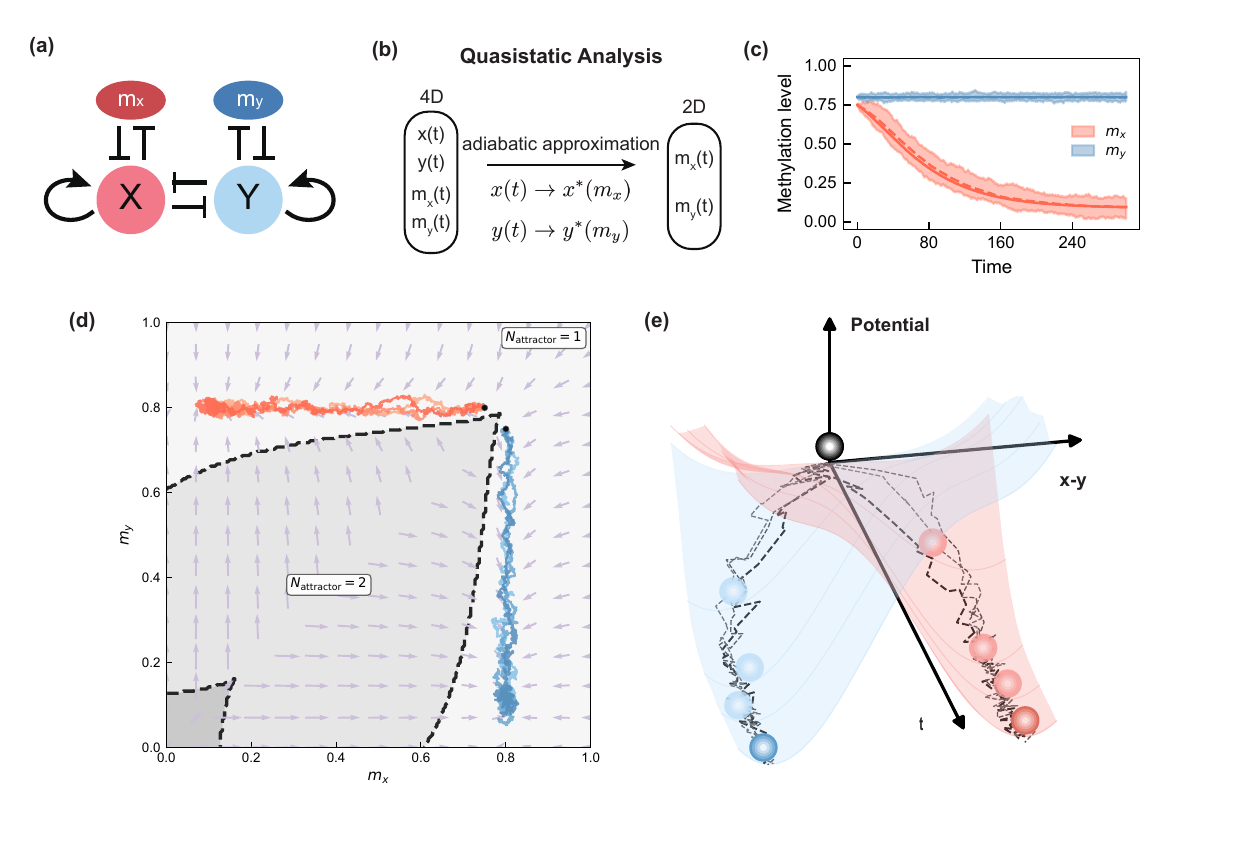}
  \caption{\textbf{A methylation-coupled fate toggle commits by single-well drift.}
    \textbf{(a)}~Two self-activating, mutually repressing genes each carry a local methylation
    state.
    \textbf{(b)}~Time-scale separation reduces the four-dimensional system to slow motion in
    methylation space $(m_x,m_y)$.
    \textbf{(c)}~Reduced trajectories agree with ensemble means from the full stochastic model.
    \textbf{(d)}~Attractor-count map for the frozen fast subsystem. Representative trajectories
    remain in the single-attractor region, showing commitment by single-well drift.
    \textbf{(e)}~The effective expression potential moves as methylation asymmetry grows,
    displacing the preferred attractor.}
  \label{fig:fig3}
\end{figure*}

In the single-locus model, $x$ is imposed from outside.
The next step is to let the variable that reads the promoter be produced by the circuit itself.
The minimal closure is positive autoregulation: the gene product $x$ activates its own production
through the same promoter response, whose threshold depends on methylation
(Fig.~\ref{fig:fig2}a).
The expression dynamics are
\begin{equation}
  \frac{dx}{dt} = \frac{x^n}{x^n + K^n(m)} - \gamma x + \xi_x,
  \label{eq:self-activation-expression}
\end{equation}
with promoter methylation following the aforementioned dynamics, Eq.~\eqref{eq:methylation-dynamics}.
Figure~\ref{fig:fig2}b shows the flow field of this two-dimensional dynamical system.
The arrows are nearly horizontal over most of the $(x,m)$ plane, consistent with rapid
transcriptional response and slow methylation turnover
\cite{lannan2022epigenetic,bruno2024analysis}.
The nullclines show the complementary effect of methylation on the fast subsystem: at fixed
$m$, the circuit is an ordinary self-activating gene with a methylation-dependent activation
threshold, and changing $m$ changes the number, position, and stability of the expression fixed
points.

Such separation of time scales allows us to do a quasi-steady-state (adiabatic) approximation.
For each fixed methylation level, the fast expression dynamics relax to a stable branch
$x^\ast(m)$ (Fig.~\ref{fig:fig2}c), where $x^\ast(m)$ is a stable root of the deterministic drift
in Eq.~\eqref{eq:self-activation-expression}, namely
$\left.dx/dt\right|_{x=x^\ast(m)}=0$.
Plugging $h(x^\ast(m),m)$ into Eq.~\eqref{eq:methylation-dynamics} then gives the autonomous
methylation dynamics $m(t)$.
Figure~\ref{fig:fig2}d shows the direction and fixed points of this slow dynamics.

To represent these regulatory dynamics geometrically, we introduce an expression landscape.
At fixed methylation, the remaining fast coordinate is one-dimensional, so the deterministic
regulatory drift can be represented by an effective potential,
\begin{equation}
  -\frac{\partial U}{\partial x}
  =
  \frac{x^n}{x^n + K^n(m)} - \gamma x .
\end{equation}
Stable expression states are minima of $U(x;m)$.
This deterministic potential is the small-noise limit of a landscape description; when intrinsic
noise is large, the relevant object is instead the free-energy landscape
$U(x;m)=-\log P_{\mathrm{ss}}(x | m)$, where $P_{\mathrm{ss}}(x | m)$ is the steady-state
distribution of expression at fixed methylation \cite{kim2007potential}.

In a Waddington-style picture, landscape changes are usually driven by external developmental
signals or imposed control parameters.
Here, if methylation is ignored or held fixed, the self-activating gene has the usual static
bistable landscape: in the absence of external input, the surface does not change over time
(Fig.~\ref{fig:fig2}e).
With methylation feedback active, by contrast, the landscape $U(x;m)$ reshapes itself as the
internal variable $m$ evolves.
Different initial methylation states select different branches of this autonomous landscape
motion; along each branch, one well becomes dominant and deepens over time
(Fig.~\ref{fig:fig2}f).
Trajectories simulated from the full two-dimensional model, Eqs.~\eqref{eq:methylation-dynamics}
and \eqref{eq:self-activation-expression}, follow the moving landscape derived from the
adiabatic approximation.

\begin{figure*}[t!]
  \centering
  \includegraphics[width=\textwidth]{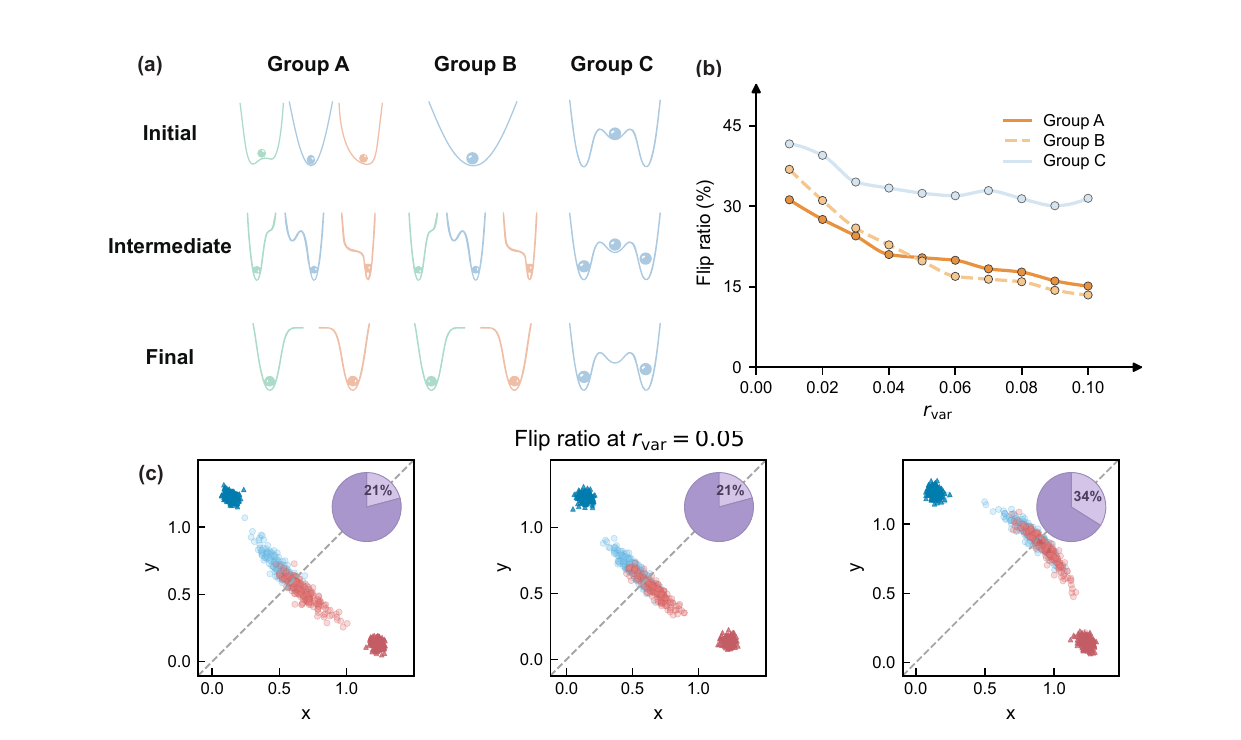}
  \caption{\textbf{Autonomous landscape evolution makes early fate bias more predictive.}
    \textbf{(a)}~Three ensemble conditions compare evolving methylation with and without initial
    methylation variance against a frozen-landscape control.
    \textbf{(b)}~Flip ratio decreases as expression variance grows, but evolving landscapes show
    fewer fate reversals than the frozen control at matched variance.
    \textbf{(c)}~At $\mathrm{Var}(x-y)=0.05$, the two evolving-landscape groups have similar flip
    ratios and both differ from the frozen-landscape group, indicating that autonomous landscape
    motion stabilizes early expression bias.}
  \label{fig:fig4}
\end{figure*}

The same feedback becomes more consequential in a fate circuit.
We embed it in a toggle switch, the minimal architecture for two mutually exclusive expression
states \cite{kim2007potential,alradhawi2019multimodality}.
The two genes, $x$ and $y$, self-activate, repress each other, and carry separate promoter
methylation states $(m_x,m_y)$ (Fig.~\ref{fig:fig3}a).
The fast expression dynamics are
\begin{align}
  \frac{dx}{dt} &= \frac{x^n}{x^n + K^n(m_x)} + \frac{k^n}{y^n + k^n}
                   - \gamma x + \xi_x, \\
  \frac{dy}{dt} &= \frac{y^n}{y^n + K^n(m_y)} + \frac{k^n}{x^n + k^n}
                   - \gamma y + \xi_y.
\end{align}
Each promoter follows the methylation dynamics in Eq.~\eqref{eq:methylation-dynamics};
$h(x,m_x)$ and $h(y,m_y)$ govern both self-activation and methylation dynamics at the
corresponding promoters, respectively.

We initialize the toggle near a symmetric, highly methylated state,
$m_x\simeq m_y\simeq m_0=0.8$, so that 80\% of CpGs are methylated at both modeled promoters.
This high-methylation starting point is consistent with genome-wide methylomes of human
embryonic stem cells, where about 80\% of CG observations are methylated
\cite{lister2009human}.
As above, we use an adiabatic approximation to derive the autonomous dynamics of the slow
variables, methylation $(m_x,m_y)$.
For each fixed $(m_x,m_y)$, we find the stable roots $(x^\ast,y^\ast)$ of the deterministic
expression dynamics in Eqs.~(6--7), then plug the corresponding occupancies $h(x^\ast,m_x)$ and
$h(y^\ast,m_y)$ into the methylation dynamics.
This gives an autonomous flow for $(m_x,m_y)$ (Fig.~\ref{fig:fig3}b).
The ensemble of simulated methylation dynamics agrees well with the prediction from the
adiabatic approximation (Fig.~\ref{fig:fig3}c).

To connect this methylation flow to fate choice, we construct the expression landscape.
In a toggle switch, the difference $x-y$ is the natural coordinate of fate asymmetry.
We therefore project the deterministic drift in Eqs.~(6--7) onto this direction and integrate the
projected force to obtain an effective landscape along $x-y$.
There exist two branches of landscapes, selected by the initial methylation state and corresponding
to the two differentiation directions.
Strikingly, along each branch the landscape contains a single well whose minimum drifts toward
one fate and deepens over time (Fig.~\ref{fig:fig3}e).

Because the landscape is fully controlled by $(m_x,m_y)$, we can classify methylation space by
the number of wells $N_{\rm wells}$ in the landscape.
Figure~\ref{fig:fig3}d marks regions with $N_{\rm wells}=1,2,$ or $3$.
Trajectories starting near $(m_0,m_0)$ remain in the $N_{\rm wells}=1$ region, showing that
commitment need not proceed through the birth of two competing fate wells.
We call this mode single-well drift: the preferred expression state moves continuously under
autonomous methylation feedback, in contrast to the bifurcation picture in which fate choice is
explained by the emergence of multistability in the landscape.

Single-well drift also changes what an early expression bias means.
In a frozen landscape, a small value of $x-y$ before clear separation is likely a reversible
fluctuation.
Along an autonomously moving landscape, the same small asymmetry can already mark the branch the
trajectory is following.
This might provide a mechanism for lineage-tracing observations in which early transcriptional
bias contains information about later fate before large expression differences are visible
\cite{wang2022cospar}.

We measure reversibility of transcriptional states by the \emph{flip ratio}: the
fraction of trajectories whose final fate has the opposite sign from their early expression bias
(Methods).
We compare evolving methylation from variable or identical initial conditions with frozen
methylation (Fig.~\ref{fig:fig4}a).
The variance of $x-y$ serves as a common clock for visible fate separation.
As this variance grows, the flip ratio decreases in all ensembles (Fig.~\ref{fig:fig4}b).
At the same variance, however, both settings with methylation dynamics have lower flip ratios than the
frozen control.
The same visible expression separation is therefore more predictive when the landscape can
autonomously reshape.

At $\mathrm{Var}(x-y)=0.05$, the two settings with methylation dynamics have similar flip ratios,
and both are notably lower than the case with frozen methylation (Fig.~\ref{fig:fig4}c).
Initial methylation heterogeneity is therefore not the main source of the effect.
The decisive ingredient is autonomous landscape evolution.
A weak expression bias becomes encoded in the slow coordinate and reappears as a regulatory
preference before large expression differences have appeared.

\section*{Discussion}

These results turn DNA methylation from a regulatory memory into a dynamical control variable.
Once promoter methylation is allowed to evolve with regulatory activity, it no longer serves only
as a record of the past.
It parameterizes the fast expression landscape and changes the regulatory conditions encountered
next.
The landscape can therefore change without an external schedule.

This view extends, rather than replaces, landscape theory.
Waddington-like landscapes have been quantified using attractors, stationary distributions, and
quasi-potentials for gene-regulatory networks
\cite{kim2007potential,bhattacharya2011deterministic,vittadello2024towards,rabajante2015branching}.
Epigenetic feedback has also been proposed to generate homeorhetic landscape motion
\cite{matsushita2020homeorhesis}, and slow promoter-state dynamics can generate multimodal
expression landscapes, including in toggle switches \cite{alradhawi2019multimodality}.
Our contribution is to derive this motion from a local promoter closure rather than impose a
time-dependent landscape.
The same model specifies both the methylation-conditioned expression landscape and the slow
dynamics that reshape it from within.

This landscape analysis reveals single-well drift as a new commitment mode.
It allows cell fate decision without the observed trajectory entering a multiwell regime.
Single-well drift should not be read as a replacement for bifurcation.
Other parameter regimes of the same toggle contain multiple wells, and external signals can still
drive conventional landscape branching.
The specific point is that a slow internal coordinate can move the preferred expression state
before the observed trajectory enters a multiwell regime.
This changes how early fate bias can be interpreted.
In a fixed landscape, weak expression asymmetry before visible separation is likely reversible.
With feedback, the same asymmetry begins to alter future regulatory conditions.
This might provide a mechanism for lineage-tracing observations in which early transcriptional
bias predicts later fate before large expression differences are apparent \cite{wang2022cospar}.

The construction is deliberately minimal.
The methylation state is represented by a mean promoter coordinate, methylation-dependent binding
is reduced to one energy parameter, and the landscapes used in the main text are deterministic
small-noise effective potentials rather than full nonequilibrium quasi-potentials of the joint
slow-fast stochastic system.
These simplifications would not change the central lesson: a molecular memory mark can become an
internal control variable for the future dynamics of gene regulation.

\bibliographystyle{apsrev4-2}
\bibliography{bib}

\clearpage
\section*{Methods}

\subsection*{Derivation of the methylation-dependent binding affinity}

We model TF--DNA binding using a grand canonical statistical mechanical framework.
A transcription factor occupies the regulatory DNA through $n$ simultaneous molecular contacts, where $n$ is the effective coordination number of the TF--DNA interaction.
In the strongly cooperative (all-or-nothing) limit, only two binding states contribute appreciably to the partition function: the unbound state (reference energy zero) and the fully bound state in which all $n$ contacts are formed, each contributing a binding energy $\varepsilon_0 > 0$.
Treating the TF-containing nucleus as a grand canonical reservoir with TF concentration $x$ (proportional to chemical activity $e^{\beta\mu}$), the partition function is
\[
\mathcal{Z} = 1 + x^n\, e^{n\beta\varepsilon_0},
\]
and the fractional saturation reads
\[
h(x) = \frac{x^n\,e^{n\beta\varepsilon_0}}{\mathcal{Z}} = \frac{x^n}{x^n + e^{-n\beta\varepsilon_0}},
\]
which is a Hill function with half-saturation constant $K^n = e^{-n\beta\varepsilon_0}$.
The Hill coefficient $n$ thus inherits its mechanistic interpretation from the coordination number of the binding event; in practice it is treated as an effective cooperativity parameter.

CpG methylation modifies binding through an additive free energy perturbation.
Each of the $n$ contacts is destabilised when CpGs within the TF footprint are methylated: at fractional methylation level $m$ (measured relative to a resting reference $m_0$), each contact contributes an extra energy $\varepsilon_1(m - m_0)$, with $\varepsilon_1 < 0$ encoding the destabilising effect.
Under the mean-field approximation that all contacts experience the same average methylation environment, the total binding free energy becomes
\[
\Delta G(m) = -n\bigl[\varepsilon_0 + \varepsilon_1(m - m_0)\bigr].
\]
Substituting $\Delta G(m)$ into the grand canonical partition function gives the methylation-dependent saturation
\[
h(x,m) = \frac{x^n}{x^n + K^n(m)},
\]
with the half-saturation constant
\[
K(m) = \exp\!\bigl(-\beta\bigl[\varepsilon_0 + \varepsilon_1(m - m_0)\bigr]\bigr).
\]
Because $\varepsilon_1 < 0$, $K(m)$ increases monotonically with $m$: higher methylation destabilises TF binding and shifts the dose--response curve to higher TF concentrations.
Throughout this work we set $\varepsilon_0 = 0$, absorbing the reference-state affinity into the TF concentration scale, and fix $n = 4$ as the effective Hill coefficient.

\subsection*{Model parameters}

All three circuits share the same biophysical framework. Parameter values are summarised in
Table~\ref{tab:parameters}; the resting methylation level in the absence of TF occupancy is
$m_0 = p_m/(p_m + r_m)$.

\begin{table*}[t]
  \centering
  \caption{Circuit parameters.}
  \label{tab:parameters}
  \small
  \begin{tabular}{llccc}
    \hline
    Symbol & Description & Memory loop & Self-tilting & Fate toggle \\
    \hline
    $n$ & Hill coefficient & 4 & 4 & 4 \\
    $\beta$ & Inverse thermal energy $(k_BT)^{-1}$ & 1 & 1 & 1 \\
    $\varepsilon_0$ & Baseline binding energy & 0 & 0 & 0 \\
    $\varepsilon_1$ & Methylation sensitivity of binding & $-2$ & $-2$ & $-1$ \\
    $r_m$ & Demethylation rate & 0.1 & 0.02 & 0.0025 \\
    $p_m$ & Methylation rate & 0.4 & 0.1 & 0.01 \\
    $m_0 = p_m/(p_m+r_m)$ & Resting methylation level & 0.8 & 0.83 & 0.8 \\
    $\gamma$ & Expression degradation rate & --- & 1.0 & 1.6 \\
    $k$ & Mutual repression threshold & --- & --- & 1.0 \\
    $\sigma$ & Expression noise amplitude & --- & 0.1 & 0.025 \\
    $\sigma_m$ & Methylation noise amplitude & --- & 0.02 & 0.001 \\
    $\Delta t$ & Integration time step & 0.1 & 0.01 & 0.01 \\
    \hline
  \end{tabular}
\end{table*}

\subsection*{Methylation memory loop}

\textbf{Model equations and parameters.}
TF binding and dissociation equilibrate on timescales of seconds to minutes---far faster than methylation turnover (hours to cell divisions)---justifying a quasi-steady-state approximation for TF occupancy:
\[
h(x,m) = \frac{x^n}{x^n + K^n(m)}.
\]
Methylation dynamics are governed by
\[
\frac{dm}{dt} = -r_m\bigl(1 + c\,h(x,m)\bigr)\,m + p_m (1-m)\bigl(1 - h(x,m)\bigr),
\]
where $r_m$ and $p_m$ are the basal demethylation and methylation rates, and $c \ge 0$ parametrizes enhanced demethylation when the locus is bound (TET recruitment upon TF occupancy).
All simulations use $c = 0$, reducing to
\[
\frac{dm}{dt} = -r_m m + p_m (1-m)\bigl(1 - h(x,m)\bigr).
\]
The resting methylation level is $m_0 = p_m/(p_m + r_m)$.
Equations are integrated numerically using the forward Euler method with time step $\Delta t = 0.1$.
Parameter values are listed in Table~\ref{tab:parameters}.

\textbf{Two-stimulus protocol.}
The inducer concentration $x(t)$ follows a piecewise-constant waveform: a first stimulus of strength $x_1 = 2$ is applied during $[0,\, T_1]$ with $T_1 = 10$; after a stimulus-free interval $\Delta T = 2$, a second stimulus of variable strength $x_2$ is applied during $[T_1 + \Delta T,\; T_1 + \Delta T + T_2]$ with $T_2 = 2$.
For the naive (unprimed) condition, the first stimulus is omitted and the system is initialised at $m_0$.

\textbf{EC50 definition.}
For each second-stimulus strength $x_2$ drawn from 100 evenly spaced values in $[0, 2]$, the response is the time-averaged TF occupancy $h(x,m)$ during the final time unit of the second stimulus.
The EC50 is the $x_2$ value at which the response equals half its population maximum, determined by linear interpolation between adjacent data points.
Memory strength is reported as the EC50 shift:
\[
\Delta\mathrm{EC50} = \mathrm{EC50}_{\rm naive} - \mathrm{EC50}_{\rm primed},
\]
which is positive when priming increases sensitivity (lowers the effective EC50).

\subsection*{Self-tilting expression landscape}

\textbf{Model equations and parameters.}
Gene expression and methylation evolve as coupled Itô stochastic differential equations:
\begin{align*}
dx &= \left(\frac{x^n}{x^n + K^n(m)} - \gamma x\right)dt + \sigma\,dW_x, \\
dm &= \left(-r_m m + p_m(1-m)\bigl(1 - h(x,m)\bigr)\right)dt + \sigma_m\,dW_m,
\end{align*}
where $dW_x$ and $dW_m$ are independent Wiener increments.
These are integrated via the Euler--Maruyama scheme ($\Delta t = 0.01$), with $x$ clipped below zero and $m$ clipped to $[0,1]$ after each step.
Parameter values are listed in Table~\ref{tab:parameters}.

\textbf{Effective potential.}
For a fixed methylation level $m$, the effective potential is obtained by numerical integration of the deterministic drift:
\[
U(x;\,m) = -\int_0^x \left(\frac{{x'}^n}{{x'}^n + K^n(m)} - \gamma x'\right) dx',
\]
evaluated on a uniform discrete grid in $x$.
This frozen-$m$ construction is used to interpret Fig.~\ref{fig:fig2}e.
Fig.~\ref{fig:fig2}b shows the deterministic skeleton (vector field, nullclines, and fixed points) in the $(x,m)$ plane.
The dynamic landscape panel Fig.~\ref{fig:fig2}f is obtained by evaluating $U(x;\,m(t))$ at successive snapshots along a representative methylation trajectory $m(t)$.

\textbf{Quasistatic reduction (Figs.~\ref{fig:fig2}c--d).}
For each $m$ on a dense grid, attractors of the fast equation (roots of the $x$-drift) are found numerically.
A single attracting branch $x^\ast(m)$ is followed by continuity in $m$ so that $x^\ast$ selects the same expression state across adjacent $m$ where multiple stable roots exist.
The reduced drift is $F(m) = -r_m m + p_m(1-m)\bigl(1 - h(x^\ast(m),m)\bigr)$, using the same $h(x,m)$ as in the full model.
Fig.~\ref{fig:fig2}d plots $F(m)$ (effective slow flow) and marks its zeros; the one-dimensional potential $-\int F(m)\,dm$ is evaluated by trapezoidal quadrature for display where needed.

\subsection*{Methylation-coupled fate toggle}

\textbf{Model equations and parameters.}
We model the fate decision process as symmetry breaking from a symmetric initial condition toward one of two stable differentiated attractors \cite{saez2022dynamical, huang2007bifurcation, ferrell2012bistability}, consistent with the established view that bipotent progenitor cells reside near a saddle point separating two fate basins.
The methylation-coupled toggle switch evolves as:
\begin{widetext}
\begin{align*}
dx &= \left(\frac{x^n}{x^n + K^n(m_x)} + \frac{k^n}{y^n + k^n} - \gamma x\right)dt + \sigma\,dW_x, \\
dy &= \left(\frac{y^n}{y^n + K^n(m_y)} + \frac{k^n}{x^n + k^n} - \gamma y\right)dt + \sigma\,dW_y, \\
dm_x &= \left(-r_m m_x + p_m(1-m_x)\bigl(1 - h(x,m_x)\bigr)\right)dt + \sigma_m\,dW_{m_x}, \\
dm_y &= \left(-r_m m_y + p_m(1-m_y)\bigl(1 - h(y,m_y)\bigr)\right)dt + \sigma_m\,dW_{m_y},
\end{align*}
\end{widetext}
with $h(x,m) = x^n/\bigl(x^n + K^n(m)\bigr)$ as in the single-gene circuits above.
Parameter values are listed in Table~\ref{tab:parameters}.

\textbf{Reduced dynamics and landscapes (Fig.~\ref{fig:fig3}b--e).}
Lineage commitment is captured by the expression asymmetry $v = (x - y)/2$, which changes sign at fate reversal, while the total expression $u = (x + y)/2$ sets the overall activity level.
Fig.~\ref{fig:fig3}b sketches the quasistatic reduction of the four-dimensional stochastic dynamics to slow methylation coordinates $(m_x,m_y)$.
Fig.~\ref{fig:fig3}c compares ensemble-mean methylation trajectories from the full model (dashed) with trajectories of the reduced dynamics (solid).
Fig.~\ref{fig:fig3}d shows the reduced vector field in $(m_x,m_y)$ together with a map of the number of stable attractors of the frozen fast $(x,y)$ subsystem (white: one attractor; lighter and darker shading: two and three attractors, respectively); overlaid trajectories illustrate differentiation within the single-attractor region.
Fig.~\ref{fig:fig3}e displays the effective expression-space potential $U(x,y)$ at successive methylation states along a representative trajectory.

For given $(m_x, m_y, u)$, the drift along $v$ is
\[
\begin{aligned}
A(v;\, m_x, m_y, u) = \frac{1}{2}\bigl[
&F(u+v,\, u-v,\, m_x) \\
&- F(u-v,\, u+v,\, m_y)\bigr],
\end{aligned}
\]
where $F(x, y, m) = x^n/(x^n + K^n(m)) + k^n/(y^n + k^n) - \gamma x$ is the deterministic drift of $x$.
The corresponding effective potential along $v$ is
\[
U(v;\, m_x, m_y, u) = -\int_0^v A(v';\, m_x, m_y, u)\, dv',
\]
evaluated by the trapezoidal rule on a uniform grid in $v$ and normalised so that $U(0) = 0$.
The dynamic landscape in Fig.~\ref{fig:fig3}e is generated along a trajectory by evaluating the expression-space effective potential $U(x,y)$ described in the figure caption at successive $(m_x(t),m_y(t))$, using the deterministic drift $F$ at frozen methylation.

\textbf{Flip ratio (Fig.~\ref{fig:fig4}).}
We simulate $N = 500$ independent trajectories for each of three conditions:
\begin{enumerate}
  \item \textit{Evolving landscape + $m_0$ variance} (Group~A): methylation dynamics fully active; $m_{x,0}$ and $m_{y,0}$ drawn independently from $\mathcal{N}(m_0,\, \sigma_{m_0}^2)$ with $\sigma_{m_0} = 0.05$.
  \item \textit{Evolving landscape} (Group~B): methylation dynamics active; $m_{x,0} = m_{y,0} = m_0 = 0.8$.
  \item \textit{Frozen landscape} (Group~C): methylation dynamics inactive ($r_m = p_m = \sigma_m = 0$); $m_x$ and $m_y$ are held fixed at $m_0 = 0.8$. To ensure a fair comparison, the baseline binding energy $\varepsilon_0$ is adjusted from $0$ to $-\ln(0.525) \approx 0.64$, so that the steady-state attractor positions of the frozen-landscape system match those of the \textit{Evolving landscape} condition. This is equivalent to fixing methylation at an effective level $m_{\rm eff} \approx 0.16$, mimicking the committed low-methylation state of the high-expression gene in the autonomously evolving model. Differences in flip ratio between conditions therefore reflect the autonomous landscape evolution itself, rather than differences in attractor location.
\end{enumerate}

All trajectories are initialised at $x_0 = y_0 = 1/\gamma$.
At each time step $T$, the population variance $\mathrm{Var}(x(T) - y(T))$ is recorded, and the instantaneous asymmetry sign $\mathrm{sign}(x(T) - y(T))$ of each trajectory is taken as its early fate bias at $T$.
After the simulation reaches steady state, each trajectory is assigned a final fate based on $\mathrm{sign}(x_{\rm final} - y_{\rm final})$.
The flip ratio at time $T$ is the fraction of trajectories whose early fate bias at $T$ disagrees with their final fate, averaged across the two sub-populations ($x(T) > y(T)$ and $x(T) < y(T)$).
For Fig.~\ref{fig:fig4}b, the flip ratio is plotted as a continuous function of $\mathrm{Var}(x(T) - y(T))$.
For Fig.~\ref{fig:fig4}c, the flip ratio is extracted at the time $T$ when the population variance first reaches $\mathrm{Var}(x - y) = 0.05$.

\subsection*{Code availability}

Code availability statement to be added.

\end{document}